\newcommand{\CLASSINPUTtoptextmargin}{0.78in}
\newcommand{\CLASSINPUTbottomtextmargin}{1.1in}
\documentclass[10pt,conference]{IEEEtran}
\IEEEoverridecommandlockouts

\usepackage{amsmath,amssymb,amsfonts,amsthm}
\usepackage[scr=euler]{mathalpha}
\usepackage{enumitem}
\usepackage{algorithm}
\usepackage{algpseudocode}
\usepackage{mathtools}
\usepackage{cleveref}
\crefname{figure}{Fig.}{Fig.}
\crefformat{equation}{(#2#1#3)}
\usepackage{caption}

\usepackage{graphicx}
\usepackage{textcomp}
\usepackage{xcolor}

\usepackage{cite}
\usepackage[acronym]{glossaries}
\usepackage{tabularx}

\def\BibTeX{{\rm B\kern-.05em{\sc i\kern-.025em b}\kern-.08em
    T\kern-.1667em\lower.7ex\hbox{E}\kern-.125emX}}

\theoremstyle{definition}

\newcommand\TS{\rule{0pt}{2.2ex}}

\newacronym{In-F}{In-F}{In-Factory}
\newacronym{6G}{6G}{sixth generation}
\newacronym{5G}{5G}{fifth generation}
\newacronym{SN}{SN}{sub-network}
\newacronym{HRLLC}{HRLLC}{hyper-reliable, low-latency communication}
\newacronym{URLLC}{URLLC}{ultra-reliable, low-latency communication}
\newacronym{MCS}{MCS}{modulation and coding scheme}
\newacronym{LA}{LA}{link adaptation}
\newacronym{CQI}{CQI}{channel quality indicator}
\newacronym{LSTM}{LSTM}{long short-term memory}
\newacronym{SINR}{SINR}{signal to interference plus noise ratio}
\newacronym{ACK}{ACK}{acknowledgement}
\newacronym{NACK}{NACK}{non-acknowledgement}
\newacronym{BLER}{BLER}{block error rate}
\newacronym{OLLA}{OLLA}{outer-loop link adaptation}
\newacronym{LOS}{LOS}{line-of-sight}
\newacronym{NLOS}{NLOS}{non-line-of-sight}
\newacronym{IIR}{IIR}{infinite impulse response}
\newacronym{eMBB}{eMBB}{enhanced mobile broadband}
\newacronym{TTI}{TTI}{transmission time interval}
\newacronym{IPV}{IPV}{interference power value}
\newacronym{PDF}{PDF}{probability density function}
\newacronym{MO}{MO}{modulation order}
\newacronym{CR}{CR}{coding rate}
\newacronym{ECDF}{ECDF}{Empirical cumulative distribution function}
\newacronym{EKF}{EKF}{extended Kalman filter}
\newacronym{SA}{SA}{sensor-actuator}
\newacronym{EESM}{EESM}{exponential effective signal-to noise-ratio mapping}
\newacronym{ESM}{ESM}{effective SNR mapping}
\newacronym{MIESM}{MIESM}{mutual information effective signal-to noise-ratio mapping}
\newacronym{CSI-RS}{CSI-RS}{channel state information-reference Signal}
\newacronym{MSE}{MSE}{mean square error}
\newacronym{IM}{IM}{Interference management}
\newacronym{AP}{AP}{access point}
\newacronym{UE}{UE}{user equipment}
\newacronym{CSI}{CSI}{channel state information}
\newacronym{MAC}{MAC}{medium access control}
\newacronym{INR}{INR}{Interference-to-noise-ratio}
\newacronym{3GPP}{3GPP}{third generation partnership project}
\newacronym{RDMM}{RDMM}{random directional mobility model}
\newacronym{InF-DL}{InF-DL}{indoor factory with dense clutter and low base station}
\newacronym{SNR}{SNR}{signal to noise ratio}
\newacronym{DSSM}{DSSM}{dynamic state space model}
\newacronym{DL}{DL}{downlink}
\newacronym{TDD}{TDD}{time division duplexing}
\newacronym{eSNR}{eSNR}{effective signal-to-noise ratio}
\newacronym{pdf}{pdf}{probabilty density function}
\newacronym{RRM}{RRM}{radio resource management}
\newacronym{RAE}{RAE}{relative absolute error}

\title{Dynamic Interference Prediction for In-X 6G Sub-networks}

\author{\IEEEauthorblockN{Pramesh Gautam\IEEEauthorrefmark{1}, Ravi Sharan B A G\IEEEauthorrefmark{2}, Paolo Baracca\IEEEauthorrefmark{3}, Carsten Bockelmann\IEEEauthorrefmark{1}, Thorsten Wild\IEEEauthorrefmark{2}, Armin Dekorsy\IEEEauthorrefmark{1}}
	\IEEEauthorblockA{\IEEEauthorrefmark{1}Department of Communications Engineering, University of Bremen,  Germany,\\ Email: \{gautam, bockelmann, dekorsy\}@ant.uni-bremen.de}
    \IEEEauthorblockA{\IEEEauthorrefmark{2}Nokia Bell Labs Stuttgart, Germany, Email: \{ravi.sharan, thorsten.wild\}@nokia-bell-labs.com}
    \IEEEauthorblockA{\IEEEauthorrefmark{3} Nokia Standards, Munich, Germany, Email: paolo.baracca@nokia.com}
\thanks{
		This work is supported by the German Ministry of Education and Research~(BMBF) under the grants of 16KISK109~(6G-ANNA) and 16KISK016~(Open6GHub).
}%
}
\begin{document}
\maketitle

\begin{abstract}
The sixth generation (6G) industrial Sub-networks (SNs) face several challenges in meeting extreme latency and reliability requirements in the order of $0.1-1$ ms and $99.999\text{-to-}99.99999$ percentile, respectively. Interference management (IM) plays an integral role in addressing these requirements, especially in ultra-dense SN environments with rapidly varying interference induced by channel characteristics, mobility and resource limitations. In general, IM can be achieved using resource allocation and \textit{accurate} Link adaptation (LA). In this work, we focus on the latter, where we first model interference at SN devices using the spatially consistent 3GPP channel model. Following this, we present a discrete-time dynamic state space model (DSSM) at a SN access point (AP), where interference power values (IPVs) are modeled as latent variables incorporating underlying modeling errors as well as transmission/protocol delays. Necessary approximations are then presented to simplify the DSSM and to efficiently employ the extended Kalman filter (EKF) for interference prediction. Unlike baseline methods, our proposed approach predicts IPVs solely based on the channel quality indicator (CQI) reports available at the SN AP at every transmission time interval (TTI). Numerical results demonstrate that our proposed approach clearly outperforms the conventional baseline. Furthermore, we also show that despite predicting with limited information, our proposed approach consistently achieves a comparable performance w.r.t the off-the-shelf supervised learning based baseline.

\end{abstract}

\section{Introduction}
\label{sec:introduction}

There is an increasing demand to integrate emerging vertical industry use cases as part of the future \gls{6G} communication infrastructure. The ``Network-of-Networks" concept has been identified as a potential solution, where a vertical can be integrated as a \gls{SN}~\cite{hoffmann2023secure}. A typical \gls{SN} consists of an \gls{AP} and several \glspl{UE} either connected to the \gls{AP} or communicating among themselves. Moreover, these \glspl{SN} are envisaged to operate either in a fully-autonomous fashion or under the purview of a larger network. We refer to~\cite{berardinelli2021extreme} for example \gls{SN} use-cases and deployment scenarios.

\gls{IM} has been identified as one of the major technical challenges involved in real-time \gls{SN} deployments. From an air-interface perspective, \gls{IM} can be achieved through a combination of resource allocation and \gls{LA}. While the former focuses more on appropriate radio resource management, \gls{IM} through \gls{LA} involves exploiting the underlying transmission link characteristics. In this work, we focus on improving the \gls{LA} aspects of \gls{IM} for \textit{In-X \glspl{SN}}. The In-X \glspl{SN} are a special category of \glspl{SN} targeting short-range communications with \gls{HRLLC} traffic with latency and reliability requirements ranging between $0.1-1$ ms and $99.999\text{-to-}99.99999$ percentile, respectively~\cite{recommendation2023framework}. 

A straight-forward \gls{LA} scheme used for existing \gls{URLLC} traffic involves \gls{MCS} selection based on the \gls{CQI} report and \gls{ACK}/\gls{NACK} feedback. Here, \gls{CQI} report is a quantized information characterizing the transmission link and the effective interference perceived at a \gls{UE}. Such a scheme is not readily applicable for \gls{IM} in \glspl{SN} for the following reasons. Firstly, short packets of \gls{HRLLC} trafﬁc occupy only a fraction of the physical resources in a \gls{TTI} compared to \gls{URLLC} traffic. This results in a rapid variation of interference among \glspl{SN}, which can be particularly challenging to track in real-time. Added to this, the delay involved in the \gls{CQI} reporting mechanism~\cite{ramezani2021cqi} only makes it worse. Secondly, extreme low latency requirements of \glspl{SN} impose stringent limitations on packet retransmissions~\textemdash~oftentimes limiting it to only one retransmission. Such requirements cannot be met with the \gls{LA} scheme described above due to ultra-dense \gls{SN} deployments. 

Several inference tools have been investigated in the literature for improved interference representation in the \gls{CQI} report and subsequently in the \gls{LA} schemes. These methods can be classified as either \gls{UE}-side or \gls{AP}-side improvements. In~\cite{pocovi2018joint}, \gls{UE}-side \gls{CQI} adjustments are proposed, where low-pass filtered interference estimates are factored into \gls{SINR} estimates. Similarly, the authors in~\cite{brighente2020interference} propose several \gls{UE}-side Gaussian kernel-based interference prediction methods. A drawback of~\cite{pocovi2018joint} and~\cite{brighente2020interference} is that these works assume that the channel estimates of interfering devices are readily available at the \glspl{UE}. The works in~\cite{pramesh2024int} and~\cite{gautam2023co} demonstrate a superior prediction error performance by using a \gls{LSTM} and federated-learning based \gls{UE}-side interference prediction for an In-X \gls{SN} scenario with deterministic traffic. While the explicit need for channel estimates is relaxed in ~\cite{pramesh2024int} and~\cite{gautam2023co}, labels or ground-truth \glspl{IPV} used for training the neural networks can be hard to obtain in reality due to the additional communication overhead in ultra-dense scenarios. In general, \gls{UE}-side methods improve the interference representation in the \gls{CQI} report; however they still do not account for transmission and protocol delays involved. Moreover, the battery limitations usually pose a computational bottleneck for the \gls{UE}-side approaches. 

For the \gls{AP}-side methods, the authors in~\cite{mahmood2020predictive} model the \glspl{IPV} using a discrete-time Markov chain and use a moving-average filter based predictor for improving the resource allocation aspects of \gls{IM}. While not primarily an interference prediction mechanism,~\cite{xu2013improving} proposes two \gls{CQI} prediction methods based on Wiener filter and cubic spline extrapolation, while accounting for several modeling errors involved in obtaining the \gls{CQI} report. A common drawback of the existing literature is that interference/\gls{SINR} prediction mechanisms only consider the tail probabilities of the underlying \gls{pdf}. This can be challenging to formally characterize \glspl{IPV} correlated across time and oftentimes lead to an oversimplification of the \gls{LA} problem.

In this work, we propose an \gls{AP}-side low-complexity online interference prediction method performed w.r.t the \gls{SN} data transmissions. More specifically, we employ an \gls{EKF}-based predictor to obtain \gls{IPV} estimates from the \gls{CQI} report to ultimately improve the \gls{LA}. To the best of our knowledge, \gls{AP}-side methods exploiting the inter-dependencies between the \gls{CQI} report and \glspl{IPV} have not been explored in either \gls{SN} or \gls{HRLLC} literature before. The main contributions of this work are summarized below:
\begin{enumerate}[label=\arabic*)]
    \item We formally characterize the inter-\gls{SN} interference following the spatially consistent \gls{3GPP} channel model and an ON-OFF traffic model.
    \item We present a discrete-time Gaussian \gls{DSSM}, where the \glspl{IPV} are modeled as hidden/latent state with \gls{CQI} report as the corresponding observations. We also present the relevant simplifications of the \gls{DSSM} to efficiently employ the \gls{EKF}. An advantage in using the \gls{DSSM} is that the range of \glspl{IPV} is not just restricted to tail \gls{pdf}.
    \item Finally, we corroborate our proposed approach with the help of numerical results. We show that the proposed method can achieve a comparable performance w.r.t the Genie \gls{LA} baseline. In contrast to other methods it is performed solely based on the available \gls{CQI} reports as observations and without any explicit feedback/labels.
\end{enumerate}
\section{System Model}
\label{sec:systemmodel}

\begin{figure}
    \centering
    \includegraphics[width=0.48\textwidth]{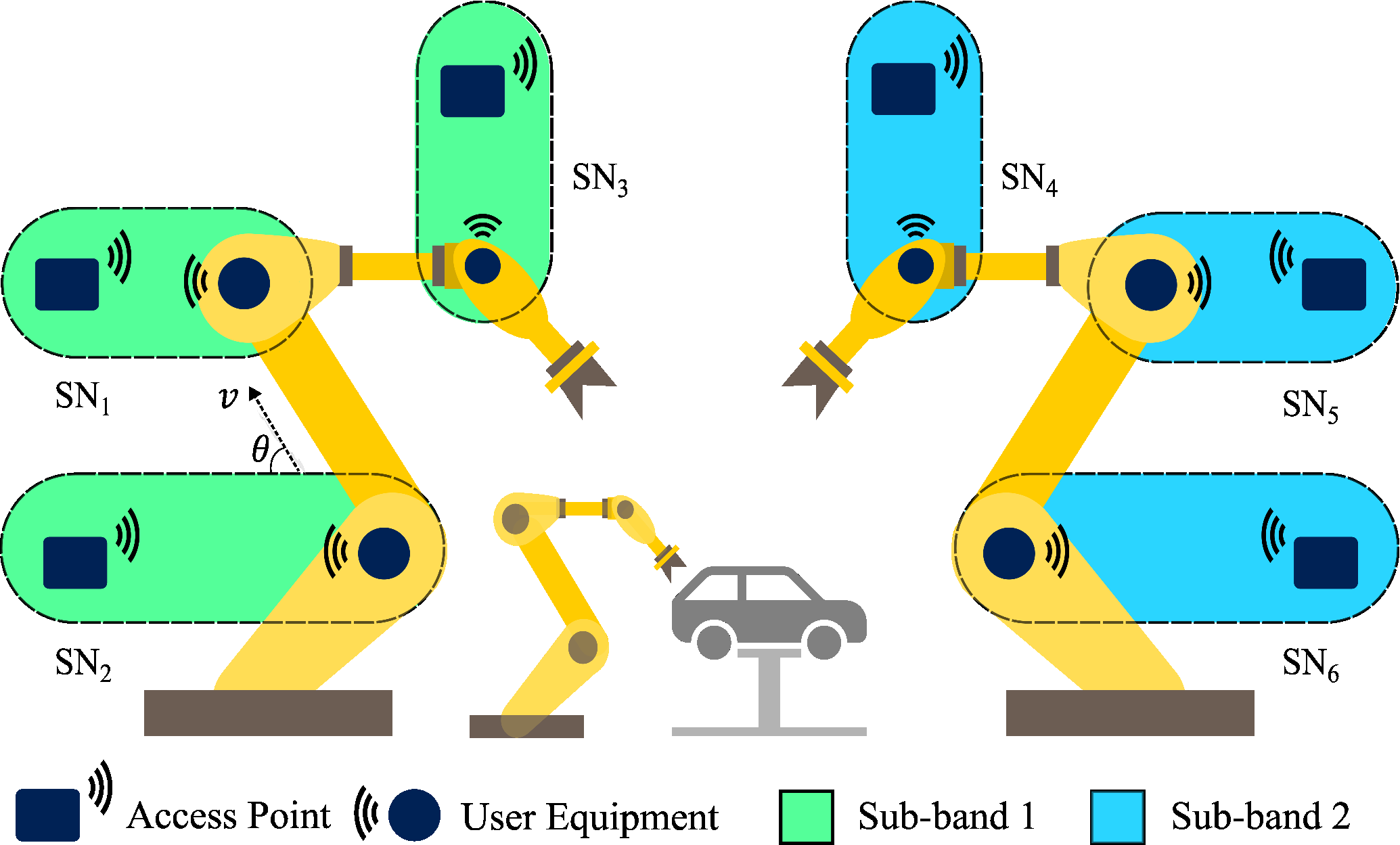}
    \caption{Example In-X \gls{SN} scenario in an automobile assembly unit. Here, several \glspl{AP} communicate with associated sensor-actuator pairs acting as \glspl{UE}.}
    \label{fig:subnet_layout}
\end{figure}

Let $\mathscr{N} = \{1,2,\ldots,N\},~|\mathscr{N}|=N$, denote the set of In-X \glspl{SN} distributed according to a uniform random variable in a confined area $\mathscr{A}$. Each \gls{SN} $n \in \mathscr{N}$ consists of a single \gls{AP} serving a single 
 \gls{UE}
 % $\mathscr{M} = \{1,2,\ldots,M\}, |\mathscr{M}|=M$ \glspl{UE}
 in the \gls{DL}\footnote{Recent works~\cite{bagherinejad2024comparative,adeogun2022multi} also consider \gls{SN} scenarios with one \gls{UE} per \gls{AP}.}. Since an \gls{AP} uniquely identifies a \gls{SN}, the terms \gls{SN} and \gls{AP} are used interchangeably in this paper. Each \gls{SN} is assumed to be moving with constant velocity $v$ with the direction defined by $\theta \in [0,2\pi]$ (see \cref{fig:subnet_layout}). Moreover, \glspl{SN} are configured to maintain a minimum separation distance.

We assume a slotted time operation where each slot or a \gls{TTI} is denoted by $t \in \mathbb{N}$. Furthermore, the slot duration is assumed to be $\tau$ for all the $N$ \glspl{SN} w.l.o.g. The total available bandwidth $B$ is partitioned into $K,~K <\!< N$ sub-bands and a \gls{SN} is assumed to operate in the TDD mode. Furthermore, an \gls{SN} is assigned to only a single sub-band at any given \gls{TTI} $t$; however, multiple \glspl{SN} can be allocated to the same sub-band. This results in an inter-\gls{SN} interference scenario.
% Moreover, due the short-range operations of the \glspl{SN}, we assume that all \glspl{UE} in a \gls{SN} experience \textit{same} interference. This is a reasonable assumption for In-body/In-robot \gls{SN} scenarios.
On the other hand, we assume that there is no intra-\gls{SN} interference. This can be achieved by using a scheduler at the \gls{AP}, which schedules the \glspl{UE} in a near-orthogonal fashion~\cite{skapoor2018distributedscheduling}. Finally, we model the \gls{SN} traffic according to an i.i.d Bernoulli random variable w.l.o.g, i.e., a \gls{AP} either has a packet to transmit or remains idle during a \gls{TTI} $t$.

Following operations ensue in a \gls{SN} to successfully perform \gls{LA} for \gls{DL} transmissions. The \gls{AP} transmits reference signals, which are used to estimate the \gls{SINR} by the associated \glspl{UE} either for all $K$ or $\Tilde{K} <\! K$ sub-bands. The estimated \gls{SINR} is converted to \gls{eSNR} using an \gls{ESM} mechanism~\cite{pocovi2017mac}. The \gls{eSNR} value is used to prepare the \gls{CQI} report using predefined look-up tables, which is then sent to the \gls{AP} in a periodic/aperiodic manner. At the \gls{AP}, the received \gls{CQI} is mapped to the corresponding \gls{SINR} estimate using a look-up table before performing \gls{LA}. The \gls{SINR} estimate of a \gls{UE}, denoted $\hat{\gamma}_{n}$, can be written as:
\begin{align}
\label{eq:sinr}
&\hat{\gamma}_{n}(t) := \frac{S_{n}(t)}{I_{n}(t)+\sigma_w^2},
\end{align}
where, $S_{n}(t)$ denotes the signal power, which is a function of the wireless channel in the 
% $m^{\text{th}}$ 
$n^{\text{th}}$ \gls{SN}. The quantity $\sigma_w^2$ is the receiver noise power, which is assumed to be same for all \glspl{UE} w.l.o.g. Finally, $I_n(t)$ is the inter-\gls{SN} \gls{IPV} perceived by the $N^{\text{th}}$ \gls{AP}.
% $m^{\text{th}}$ \gls{UE}.
% Notice the difference in subscripts of $S_{n}(t)$ in the numerator and $I_n(t)$ in the denominator. This is possible due to the assumptions on the interference presented above.

In this work, we model $I_n(t)$ based on the spatially consistent \gls{3GPP} channel model ~\cite{3gpp2020study}, and formally express it as:
\begin{multline}
\label{eq:interferencemodeling}
I_{n}(t) := \sum_{n^{\prime} \in \hat{\mathscr{N}}(t)\setminus\{n\}}  \chi_{n^{\prime}}(t) \cdot P_{n^{\prime}}(t) \cdot \biggl(\beta |H_{\text{LOS},n^{\prime}}(t)|^2+ \\ \sqrt{(1- \beta^2)} |H_{\text{NLOS},n^{\prime}}(t)|^2\biggr),
\end{multline}
where, $\mathscr{\hat{N}}(t)\subset\mathscr{N}$ denotes the set of all \glspl{SN} operating in the same sub-band at \gls{TTI} $t$. The quantity $\chi_{n^{\prime}}(t) \sim Ber(\rho)$ denotes the ON-OFF traffic parameterized by probability $\rho$. The term $P_{n^{\prime}}(t)$ denotes the transmit power of the interfering \glspl{SN}. Furthermore, $H_{\text{LOS},n^{\prime}}(t) := h_{\text{LOS},n^{\prime}}(t)\cdot l_{\text{LOS},n^{\prime}}(t)\cdot\zeta_{\text{LOS},n^{\prime}}(t)$ is the consolidated \gls{LOS} channel where $h_{LOS,n^{\prime}}$ is the small-scale fading co-efficient correlated across \glspl{TTI} and following Rician \gls{pdf} \cite{xiao2006novel}; $l_{\text{LOS},n^{\prime}}$ is the path loss component \cite{3gpp2020study} and $\zeta_{\text{LOS},n^{\prime}}$ represents the spatially correlated shadowing effect, which follows a log-normal \gls{pdf} \cite{lu2015effects}. The value $H_{\text{NLOS},n^{\prime}}$ is the consolidated \gls{NLOS} channel defined in a way similar to the \gls{LOS} channel. The correlated small scale fading across \gls{TTI} with rayleigh \gls{pdf} is denoted as $h_{\text{NLOS},n^{\prime}}$, alongside path loss $l_{\text{NLOS},n^{\prime}}$ and shadowing $\zeta_{\text{NLOS},n^{\prime}}$ with respective \gls{NLOS} parameters.
% However, $h_{NLOS,n^{\prime}}$ follows a Rayleigh p.d.f, with the p.d.f of $l_{NLOS,n^{\prime}}$ and $\zeta_{NLOS,n^{\prime}}$ remaining the same as \gls{LOS} channel. 
The parameter $\beta \in (0,1)$ is a smoothing factor which prevents abrupt fluctuations in the channel response caused by transitions between \gls{LOS} and \gls{NLOS} conditions \cite{3gpp2020study}. 
%Furthermore, $\beta$ is parameterized w.r.t the minimum separation distance, $d_{\text{sep}}$.
With the system model in place, we present the problem description in the following section.

\section{Link Adaptation Problem}
\label{sec.probformulation}

The \gls{LA} problem involves appropriate \gls{MCS} selection subject to reliability constraints. As mentioned in section~\ref{sec:systemmodel}, an \gls{AP} performs \gls{MCS} selection based on the \gls{CQI} report received from its associated \gls{UE}. Formally, the \gls{LA} problem for the \gls{UE} of the $n^{\text{th}}$ \gls{SN} at \gls{TTI} $t$ can be expressed as:
\begin{align}
\label{eq:laprob}
\lambda_{n}^*(t)|\hat{\gamma}_{n}(t) = \sup\{R(\lambda_{n}(t))|\varepsilon(\lambda_{n}(t)) \leq \Bar{\varepsilon}\},
\end{align}
where, $\lambda_{n}^*(t) \in \mathscr{L}$ is the optimal \gls{MCS} value with $\mathscr{L}, |\mathscr{L}|=L$ being the set of all feasible \gls{MCS} values. The quantity $R(.)$ denotes the achievable ergodic rate of 
%the  $m^{\text{th}}$ 
\gls{UE} of $n^{\text{th}}$ \gls{AP}. Finally $\varepsilon(t)$ and $\Bar{\varepsilon}$ represent the achieved \gls{BLER} and target \gls{BLER} values, respectively.

Following practical challenges arise while solving~\eqref{eq:laprob}:
\begin{enumerate}[label=(\alph*)]
    \item The interference experienced by \glspl{UE} is not sufficiently reflected in~\eqref{eq:laprob} due to both compression losses involved in \gls{ESM} and the limited quantization levels collectively characterizing the \gls{CQI} report.
    \item Secondly, an \gls{AP} is more likely to perform \gls{MCS} selection based on outdated \gls{CQI} and subsequently the interference information due to inherent protocol overhead and transmission delays involved.
\end{enumerate}

While (a) and (b) are not just specific to \gls{LA} for \glspl{SN}, not accounting for them in~\eqref{eq:laprob} can have a detrimental effect on the overall latency and reliability requirements, especially in \glspl{SN}. A straightforward way to include interference information in~\eqref{eq:laprob} is by explicitly estimating the \glspl{IPV} at an \gls{UE} and sending them back to the associated \gls{AP}. However, this is impractical in ultra dense \gls{SN} scenarios since it introduces additional signalling and computational overhead. In this work, we resort to an online signalling-free interference prediction scheme by exploiting the \gls{CQI} available at the \gls{AP}, which we describe in the next section.

\section{State Space Modeling}
\label{sec:dssmcharacterization}

Prior to discussing the proposed interference prediction scheme, we first present the relevant \gls{DSSM} of \glspl{IPV} modeled at the \gls{AP} w.r.t the \gls{CQI} reports of the associated \glspl{UE}. A \gls{DSSM} deals with time-dependent problems involving latent variables which effectively describes the evolution in the state of the underlying system. Typically, these latent variables are associated with a set of observations. In this work, we model the \glspl{IPV} as the state/latent variables of the \gls{DSSM}. This is possible since the \gls{CQI} report implicitly captures the interference information, even when it is subjected to protocol delays, compression losses and modeling perturbations. Secondly, \glspl{IPV} demonstrate correlation across \glspl{TTI} due to the underlying channel model as shown in~\eqref{eq:interferencemodeling}. This is helpful in characterizing the state/latent variable transitions. Given this context, we model a discrete-time Gaussian \gls{DSSM} at the $n^{\text{th}}$ \gls{AP}, which is formally expressed as follows:
\begin{subequations}
\label{eq:dssm}
\begin{align}
    &I_n(t) := F(I_n(t-1)) + \nu(t),&\nu(t) &\sim \mathcal{N}(0,\sigma^2_F) \label{eq:dssma} \\
    &Y_{n}(t) := G(I_n(t)) + u(t),&u(t) &\sim  \mathcal{N}(0,\sigma^2_G). \label{eq:dssmb}
\end{align}
\end{subequations}
Here,~\eqref{eq:dssma} represents a process model describing the dynamic evolution of the \gls{IPV} over the \glspl{TTI} modeled as the state/latent variable of the \gls{DSSM}. Similarly,~\eqref{eq:dssmb} corresponds to a measurement model relating the \gls{CQI} report of the associated \gls{UE}, $Y_{n}(t)$ to the \gls{IPV}, $I_n(t)$. The function $F(.)$ denotes the \gls{IPV} transitions from one \gls{TTI} to another, whereas $G(.)$ is the observation function represented by \gls{SINR}~\eqref{eq:sinr}. The quantity $\nu(t)$ models the process noise in~\eqref{eq:dssmb}. Similarly, $u(t)$ corresponds to the measurement noise which is jointly characterized w.r.t the error due to compression losses of \gls{ESM} and \gls{CQI} quantization. Both $\nu(t)$ and $u(t)$ are assumed to follow a Gaussian \gls{pdf} with variance $\sigma^2_F$ and $\sigma^2_G$, respectively. Although the \gls{CQI} reporting delay is considered to be one \gls{TTI} in~\eqref{eq:dssm}, it can be easily accounted for with minimal modifications to~\eqref{eq:dssma}. Given this \gls{DSSM}, we are interested in mapping $Y_{n}(t)$ to an estimate of \gls{IPV} i.e., $\hat{I}_n(t)$ at every $t$ in an online fashion~\textemdash~a process usually referred to as \textit{filtering} or \textit{online state recovery} in the \gls{DSSM} literature.
\section{Extended Kalman Filter Approximations}
\label{sec:ekfprediction}

\gls{EKF} is a popular Kalman Filter heuristic used for solving the non-linear filtering problem of a \gls{DSSM}. The \gls{EKF}, at every prediction instance, provides an instantaneous state estimate based on the observations. To achieve this, the \gls{EKF} operates on linear approximations of the functions characterizing transition dynamics and/or the observations. In this section, we present the necessary approximations to~\eqref{eq:dssm} such that the \gls{EKF} can be efficiently used for \gls{IPV} prediction, which is in-turn used for \gls{LA} improvements. To this extent, we first exploit the correlated nature of small-scale fading coefficients to approximate the transition dynamics of $F(.)$ in~\eqref{eq:dssma}. More specifically, we identify that the correlation factor of $h_{NLOS,n},~\forall n \in \hat{\mathscr{N}}(t)$ in~\eqref{eq:interferencemodeling} can be approximated using a zeroth-order Bessel function, $B_{0}(\omega_{n} \cdot \tau)$~\cite{xiao2006novel}. Here, $\omega_n = 2 \pi f_{d_n}$ denotes the Doppler spread, with $f_{d_n}$ being the Doppler frequency and $\tau$ being the \gls{TTI} duration of the interfering \glspl{SN} as described in section~\ref{sec:systemmodel}. Based on this, we approximate the correlation factor of \glspl{IPV}, $\alpha_{F_n} \in (0,1)$ as follows:
\begin{align}
\label{eq:alphaF}
    \alpha_{F_{n}} = \frac{1}{N}\sum_{\mathscr{N}} B_{0}(\omega_n \cdot \tau).
\end{align}

Additionally, instead of simply using the previous estimate $\hat{I}_n(t-1)$, we reinforce the \gls{EKF} prediction by adopting the correlation matching criterion \cite{ghandour2012use}. Formally, this can be represented as a convex combination of the past two predicted \glspl{IPV}, i.e., $\hat{I}_n(t-1)$ and $\hat{I}_n(t-2)$. With a slight abuse of notation in the \gls{TTI} index representation, we denote the consolidated previous \gls{IPV} estimate as $\Tilde{I}_n(t-1)$. In summary, the linear approximation of the transition dynamics $\Tilde{F}(.)$, can be expressed as follows:
\begin{align}
\label{eq:approxprocessfunction}
    \Tilde{F}(\Tilde{I}_n(t-1)) \approx \alpha_{F_{n}} \cdot \hat{I}_n(t-1) + (1- \alpha_{F_{n}}) \cdot \hat{I}_n(t-2).
\end{align}

On the other hand, we follow a straightforward approximation of $G(.)$ in~\eqref{eq:dssmb} by taking the partial derivative of the $\hat{\gamma}_{n}(t)$ from~\eqref{eq:sinr} w.r.t the \gls{IPV} estimate, $\hat{I}_n(t)$. In a practical situation where the function characterizing \gls{CQI}-\gls{SINR} mapping at the \gls{AP} is not available, an equivalent approximation for a \gls{UE}'s signal power, $S_{n}$ can be used in the numerator of~\eqref{eq:sinr} for an appropriate representation of $G(.)$. Moreover, the moments of the joint distribution characterizing $u(t)$ are hard to obtain in practical real-time \glspl{SN}. Thus, we approximate $u(t)$ with a product of two marginal Gaussian random variables associated with, $u_1(t)$, the error in \gls{ESM} and $u_2(t)$, the \gls{CQI} quantization function, respectively. We denote this approximated random variable as $\Tilde{u}(t)$, whose variance is given by~\cite{bromiley2003products}:
\begin{align}
\label{eq:approxvariance}
 \Tilde{\sigma}^2_{G} = \frac{\sigma^{2}_{G_{1}} \cdot \sigma^{2}_{G_{2}}}{\sigma^{2}_{G_{1}} + \sigma^{2}_{G_{2}}},
\end{align}
where $\sigma^{2}_{G_{1}}$ and $\sigma^{2}_{G_{2}}$ denotes the variance associated with $u_1(t)$ and $u_2(t)$, respectively.

\begin{figure}
    \centering
    \includegraphics[width=0.48\textwidth]{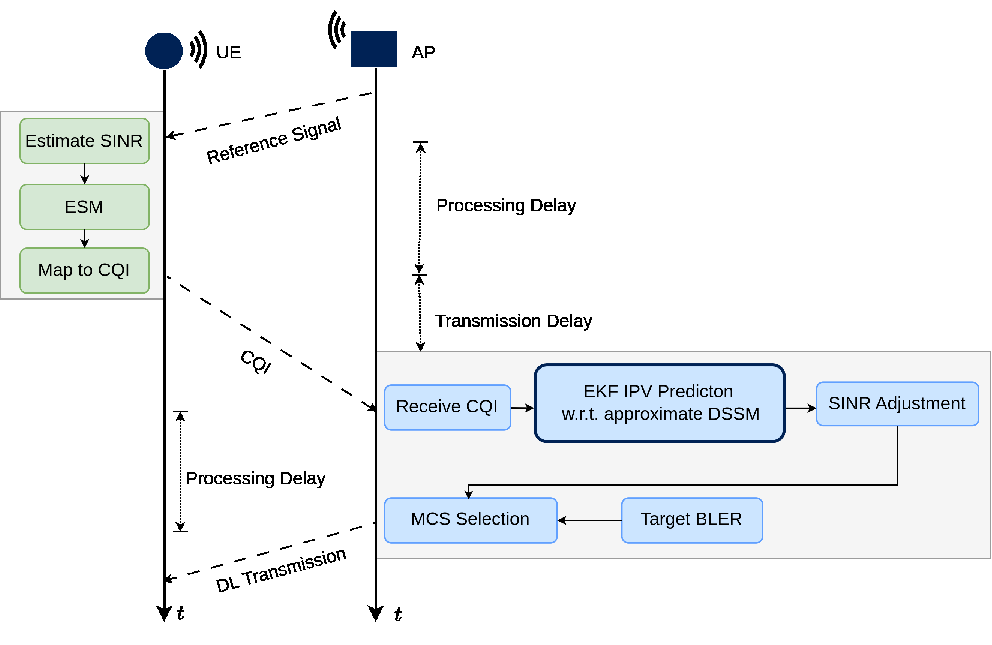}
    \caption{\gls{LA} adjustments with \gls{IPV} prediction}
    \label{fig:la_adjustments}
\end{figure}

We would like to highlight that the key novelty of this work lies in modeling the \gls{DSSM} for online prediction of \glspl{IPV} and the subsequent approximations for the \gls{DSSM} to efficiently employ the standard \gls{EKF} algorithm. Hence, we omit the algorithmic details of the \gls{EKF} algorithm for brevity. That said, given the approximations of the functions associated with process and measurement models and the corresponding noise variance, the \gls{EKF} algorithm follows a standard two-step process involving a prediction and an update step \cite{kay1993fundamentals}. The prediction step outputs an \gls{IPV} estimate and conditional variance based on $\Tilde{F}(\Tilde{I}_n(t-1))$. Similarly, in the update step, the approximate moments are propagated to rectify the \gls{IPV} estimate based on \gls{CQI} as the observations. Once the \gls{IPV} estimate is obtained, the \gls{SINR} estimate is adjusted before performing the \gls{LA}. The overall \gls{LA} operation is illustratively summarized in Fig.~\ref{fig:la_adjustments} with the proposed \gls{IPV} prediction highlighted in blue.

\subsection{Baseline Methods}
\label{sec:baseline}
Following baseline methods are considered for comparison in this work:
\begin{enumerate}[label=\arabic*)]
    \item \textit{Genie \gls{LA}:} This baseline performs \gls{MCS} selection based on the target \gls{BLER} $\Bar{\varepsilon}$ under the assumption that a \gls{UE}'s \gls{SINR} is perfectly available at the \gls{AP}. Note that Genie \gls{LA} is impractical and is considered here as an upper bound for \gls{MCS} selection~\cite{alonzo2020urllc}.
    \item \textit{Moving-average (MA) Predictor}: In this conventional baseline (used in~\cite{pocovi2018joint} and~\cite{mahmood2020predictive}), the \gls{IPV} at \gls{TTI} $t$ is obtained as a weighted-sum of \glspl{IPV} estimates from \glspl{TTI} $t-1$ and $t-2$. An important drawback of this baseline is that computing the \gls{IPV} estimates require explicit channel measurements from the interfering devices. 
    \item \textit{\gls{LSTM} Predictor:} In this supervised learning based baseline, a pre-trained \gls{LSTM} model predicts \glspl{IPV} at each $t$ based on a predefined window of previously predicted \glspl{IPV} as its input. This baseline is synonymous to a data-driven Wiener-filtering method. Refer to~\cite{gautam2023co} for the neural network architecture details.
\end{enumerate}

\section{Numerical Results}
\label{sec:simulations}

In this section, we present the performance evaluation of the proposed approach w.r.t \gls{LA} improvements and compare it with the baseline methods using numerical simulation results. All relevant simulation parameters considered in this section are summarized in Table~\ref{table:2}.

\begin{table}
\centering
\caption{Simulation Parameters}
\label{table:2}
\resizebox{0.43\textwidth}{!}{
\begin{tabular}{ll} \hline
\textbf{Parameter}                                          & \textbf{Value}~ \\ \hline
\textbf{Deployment Parameters}                                       \\ \hline
Number of \glspl{SN} , $N$                                   & $16$              \\ 
Number of \glspl{UE}, $M$                           & $1$                \\
Interfering \glspl{SN}, $|\hat{\mathscr{N}}(t)|$                           & $4$                \\
Deployment Area & $20 \times 20 ~ \text{m}^2$            \\ 
Minimum separation distance,                                     & $4$ m            \\
Mobility Model                                              & Random Directional Model   \\
Cell Radius, $r$                                    & $2$ m            \\
Velocity, $v$                                             & $2$ m/s          \\ \hline
\textbf{Channel and PHY Parameters}             \\ \hline
Carrier frequency                             & $6$ GHz         \\
Number of sub-bands, $K$                           & $4$                \\
Frequency reuse                            & $1/4$                \\
Pathloss               & \gls{3GPP} InF-DL \cite{3gpp2018study}            \\
Shadow fading std. deviation                            & $4$ dB(LOS)$~|~$$7.2$ dB(NLOS)        \\
Decorrelation distance,           & 10 m            \\
Doppler frequency, $f_{d_n}$                   &  $80~\text{Hz}$ \\
Transmit Power          &  $0$ dBW       \TS   \\
\gls{TTI} duration &  $0.1$ ms \\
Packet size & $160$ bits \\
\hline
\textbf{\gls{DSSM} and \gls{EKF} Parameters}                                     \\ \hline
\gls{CQI} type  & wideband \TS \\
\gls{MCS} reference table & Table $5.1.3.1-3$ \cite{3gpp2018physical} \\
Number of \gls{MCS} levels, $L$  & $29$ \\
\gls{SINR} min-max difference, $\Delta$ & $4.8$ \\
\gls{ESM} error variance, $ \sigma_{G_1}^{2}$          &  $\Delta^2/(12 \times L)$    \cite{easton2010fundamentals}     \\ 
\gls{CQI} mapping error variance, $\sigma_{G_2}^{2}$       & $2\times {10}^{-9}$           \\ 
Noise Factor,        & $10$ dB           \\ 
Noise Bandwidth,          & $50$ MHz           \\
Process Noise Variance, $\sigma^2_{F}$          & $0.0042$           \\
\hline
\textbf{Baseline Parameters}                                     \\ \hline
Moving-avg. smoothing factor & $0.01$  \\
\gls{LSTM} sliding window & $30$ \\
\gls{LSTM} prediction-step & $1$ \\
\gls{LSTM} training epochs & $200$ \\
\hline
\end{tabular}}
\end{table}

\begin{figure}[hbt]
    \centering
    \includegraphics[width=0.48\textwidth]{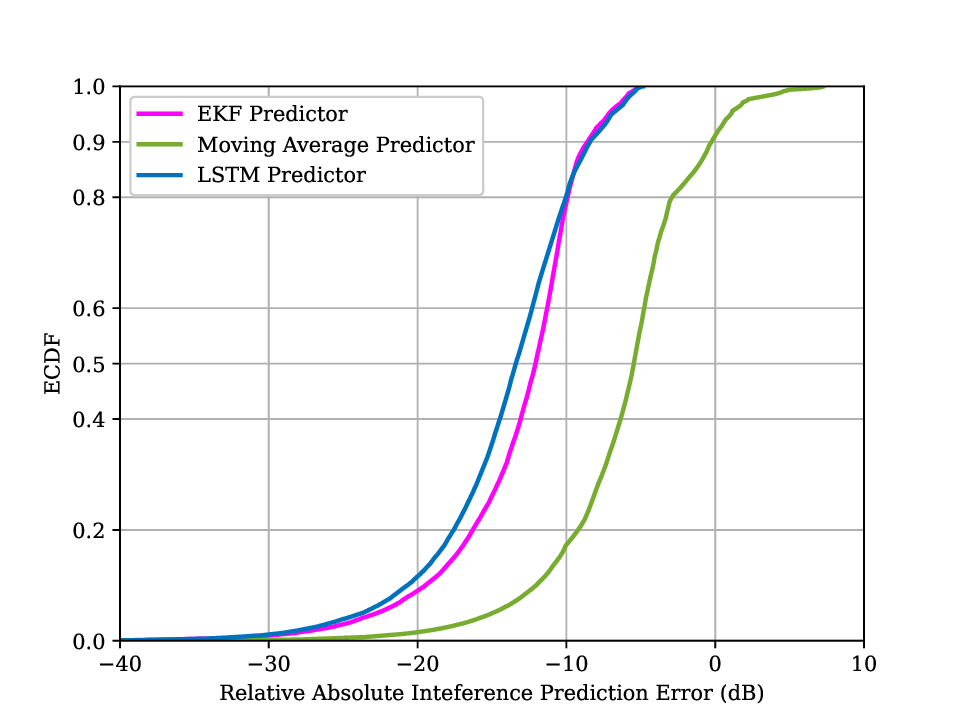}
    \caption{\gls{ECDF} of \gls{RAE}}
    \label{fig:predictiontrajectory}
\end{figure}

In Fig.~\ref{fig:predictiontrajectory}, we first compare the prediction error performance for the proposed method w.r.t the baseline methods.
% Note that the Genie predictor assumes the error-free prediction at \gls{AP}.
To do so, we use the \gls{RAE} metric defined by:
\begin{equation}\label{eq:rerr}
     \text{RAE}(t)=|(I_{n}(t)-\hat{I}_{n}(t))/I_{n}(t)|,
\end{equation}
where, $I_n(t)$ denotes the reference \gls{IPV} and $\hat{I}_n(t)$ is the predicted \gls{IPV}. 
One can observe that \gls{EKF} clearly outperforms the moving-average predictor by achieving approximately $8$dB lower prediction error. 
% This can be attributed to the fact that the moving-average predictor does not account for the underlying noise characteristics. 
This can be attributed to the fact that the moving-average predictor is not able to track the non-linear dynamics of interference accurately. One can also observe that despite being an unsupervised \gls{AP}-side method, the prediction error performance of \gls{EKF} and the \textit{pre-trained} \gls{LSTM} baseline are almost comparable. This is possible because of the most suitable approximations considered for \gls{EKF} in Section~\ref{sec:ekfprediction}.
\begin{figure}[hbt]
    \centering
    \captionsetup{justification=centering}
    \includegraphics[width=0.48\textwidth]{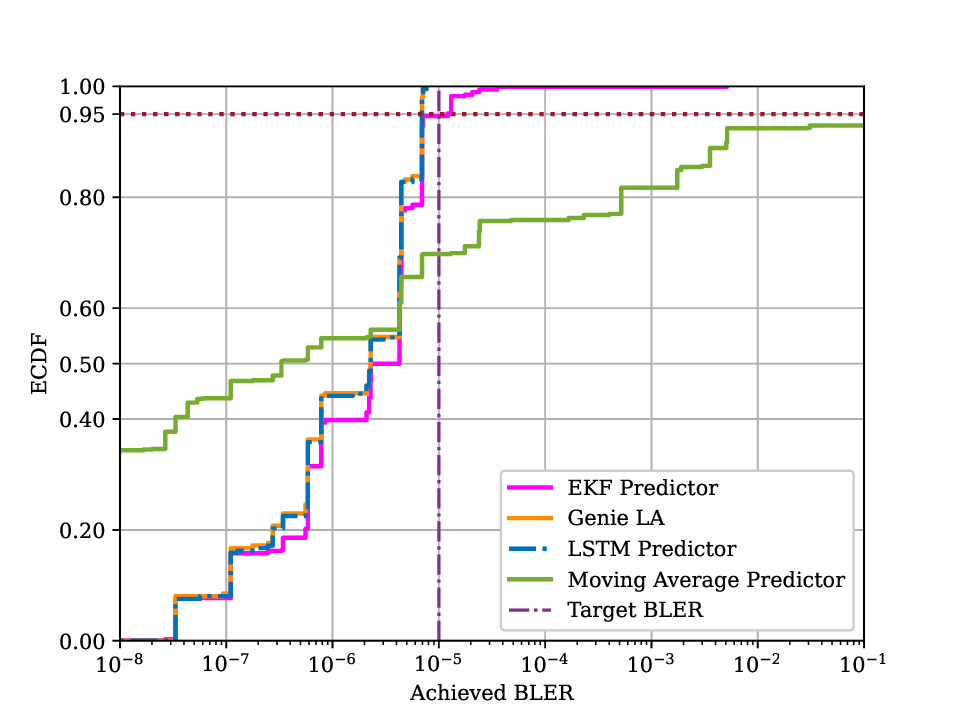}
    \caption{\gls{ECDF} of achieved BLER}
    \label{fig:bler_ecdf}
\end{figure}

\begin{figure}[hbt]
    \centering
    \includegraphics[width=0.45\textwidth]{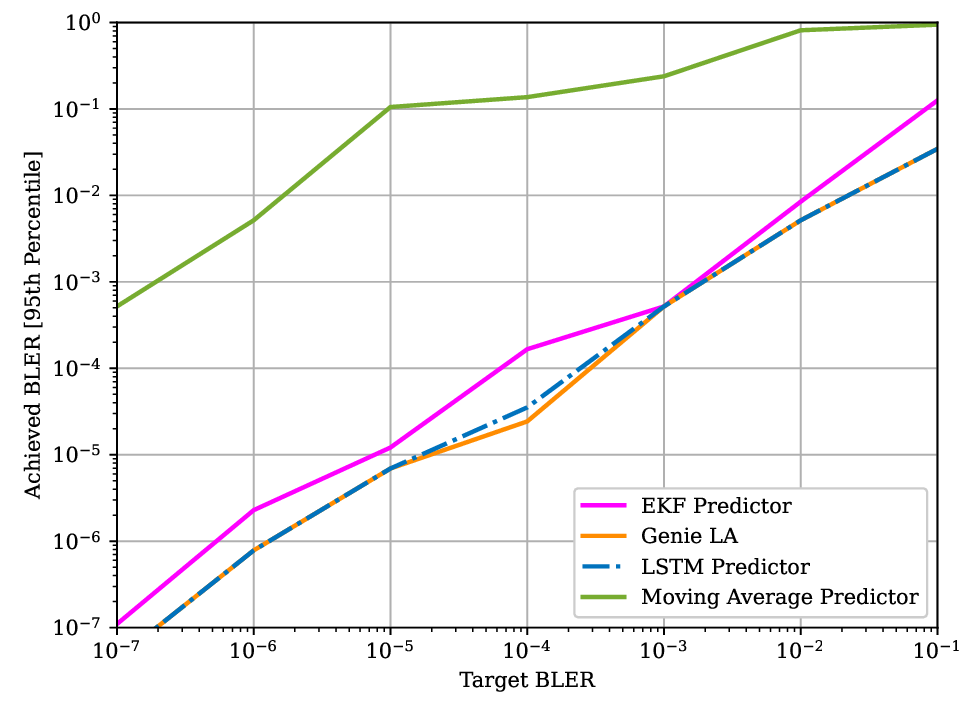}
    \caption{Achieved BLER for varying target BLER}
    \label{fig:blercomparison}
\end{figure}

In Fig.~\ref{fig:bler_ecdf} and~\ref{fig:blercomparison}, we evaluate the performance of the proposed \gls{EKF} in terms of the \gls{BLER} achieved by \gls{MCS} selection. Here, the achievable \gls{BLER} is computed based on \gls{MCS} selection and the reference \gls{SINR} lookup table mentioned in Table~\ref{table:2}. In spite of performing prediction solely based on \gls{CQI} values in an unsupervised manner, proposed \gls{EKF} achieves a target \gls{BLER} of $\Bar{\varepsilon} = 10^{-5}$ for $95\%$ of the times. Moreover, the performance of \gls{EKF} predictor strongly matches with that of \gls{LSTM} and Genie \gls{LA} baseline methods. This behaviour can be attributed to meticulous approximations w.r.t the interference modeling in~\eqref{eq:interferencemodeling} and the transition dynamics in~\eqref{eq:approxprocessfunction}.

Finally, in Fig.~\ref{fig:blercomparison}, we evaluate the \gls{EKF} predictor with respect to the Genie \gls{LA} by varying the target \gls{BLER}. Our findings indicate that as the target \gls{BLER} increases, the \gls{EKF} predictor achieves it within a slight margin of $95\%$ percentile. While the \gls{LSTM}-based predictor may offer slightly better performance, the \gls{EKF} predictor presents itself as a computationally inexpensive approach suitable for small form-factor \gls{SN} devices.
\section{Conclusions}

In this work, we propose an online and unsupervised interference prediction technique, the \gls{EKF} predictor, capable of inferring and predicting dynamic interference based on available channel information~\textemdash~specifically, the \gls{CQI} at a \gls{SN}. Despite being online and unsupervised method, the performance in terms of prediction error is comparable with \gls{LSTM} and optimal \gls{LA}, despite discrepancies in selecting variances associated with quantization, compression, and reconstruction. This suggests that with appropriate modeling and approximations, the proposed low-complexity state-space modeling has the potential to achieve a comparable performance w.r.t machine learning methods. Finally, the vector-valued \gls{DSSM} capturing multiple \glspl{UE} per a \gls{SN} \gls{AP} along with  unknown noise characteristics remain an open topic and needs further investigation.

% \bibliographystyle{IEEEtran}
% \bibliography{IEEEabrv,bibliography}

\end{document}